\documentstyle[pra,aps,multicol,graphicx,floatfig]{revtex}

\begin{document}
\draft
\title{ESR experiments in the Kondo insulator CeNiSn}
\author{S. Mair, H.--A. Krug von Nidda, M. Lohmann and A. Loidl}
\address{Experimentalphysik V, Elektronische Korrelationen und
Magnetismus, Institut f\"{u}r Physik, Universit\"{a}t Augsburg, D--86135
Augsburg, Germany}
\date{submitted to Phys.\ Rev.\ B: 6. July 1999}
\maketitle

\begin{abstract}
Below a characteristic temperature, due to hybridisation effects
Kondo insulators exhibit a gap in the electronic density of states
and behave like semiconductors. By using Gd$^{3+}$ electron spin
resonance (ESR), the compound CeNiSn was investigated as a
representative of this class. In addition, the metal--to--insulator
transition was studied as a function of doping for
CeNi$_{1-x}$Co$_x$Sn and CeNi$_{1-y}$Pt$_y$Sn. The linewidth of the
Gd resonance yields direct information about the density of states
at the Fermi energy. So the size of the gap can clearly be
estimated for the pure compound, and the closing of the gap by
substitution of Ni by Co or Pt can be followed in detail. These
results are compared to measurements of NMR, specific heat and
susceptibility.
\end{abstract}

\begin{multicols}{2}
\section{Introduction}
In recent years Ce based intermetallic compounds have been the
topic of many studies because of their large variety of ground
states, e.g. heavy--fermion behavior, non--fermi liquid behavior
near a quantum critical point, intermediate valency and a
metal--to--insulator transition in systems called Kondo insulators.
The low--temperature physics of these systems is determined by the
hybridisation of the localized Ce--4f states with the conduction
band via exchange interactions (Kondo effect). In heavy--fermion
compounds, where the 4f ions constitute a sublattice of the
crystal, this gives rise to spin fluctuations of the Ce--4f
moments, which are completely screened below a characteristic
temperature $4 \text{K} \lesssim T^* \lesssim 50 \text{K}$. Above
this temperature these compounds reveal Curie--Weiss behavior,
characteristic of local--moment paramagnets. Below $T^*$ they behave
like non--magnetic metals with strongly enhanced effective electron
masses. With increasing hybridisation ($50K \lesssim T^* \lesssim
500K$) not only the spin but also the charge of the Ce--ions starts
to fluctuate, this effect is known as intermediate valency.

In all these Kondo--lattice compounds, the coherence of the
Kondo--screening causes a pseudogap within the electronic density
of states, which evokes below $T^*$ near the Fermi energy $E_F$
\cite{gre91}. Its width is of the order of $T^*$. If this gap
develops exactly at $E_F$ and if the density of states vanishes for
zero temperature in the gap, the ground state is insulating. These
so called Kondo insulators are still in the focus of recent
research activities.

One member of this class is CeNiSn with $T^* \approx 20 K$. With
decreasing temperature the electric resistance $\rho$ shows a
drastic increase below 7K. This can be explained by the opening of
an energy gap in the electronic density of states
\cite{adr96,tak96}. As described in these papers, a serious problem
of resistance measurements was the strong dependence of the results
on the sample quality, especially on the concentrations of
impurities in the required single crystals, and it seemed, that the
purest single crystals reveal metallic behavior down to the lowest
temperatures. Meanwhile, pure CeNiSn has been established as a semi
metal with a small overlap of valence and conduction bands
\cite{Ris99}. In this case the hybridisation matrix elements vanish
at appropriate places of the Brillonin zone. Therefore the density
of states exhibits a very small but finite value in the
hybridisation gap at the Fermi energy. As bulk materials always
strongly depend on stoichiometry, defects, and heterogeneities, an
experimental method was required that can measure the gap on a
microscopic scale in polycrystalline or powdered samples. Of course
we are aware that ground--state properties of many heavy--fermion
compounds sensitively depend on stoichiometry and sample--growth
conditions. We recall the ground--states of CeCu$_2$Si$_2$
\cite{ste95} and UPt$_3$ \cite{ste84}, which reveal a variety of
magnetic and superconducting states in polycrystals and single
crystals with marginal differences in the growth conditions. CeNiSn
may well be a further candidate of this class. Up to now nuclear
magnetic resonance (NMR) and neutron scattering have been used
successfully to investigate the gap in CeNiSn \cite{oha95,sat95}.
Here, we present the first electron--spin resonance investigations
in this compound. To show that the opening of a gap is an intrinsic
property of CeNiSn, we provide experimental evidence that the gap
can systematically be suppressed (closed or shifted away from
$E_F$) by substituting Co or Pt for Ni.

As the Ce spin relaxes too fast to yield any measurable ESR signal,
small amounts of Gd$^{3+}$ were doped onto the Ce place as ESR
probe. In normal metals the linewidth $\Delta H$ of the Gd$^{3+}$
ESR follows a Korringa relaxation $\Delta H \propto b T$, where $b$
directly measures the electronic density of states at the Fermi
energy. It has been shown, that the temperature dependence of the
Gd linewidth obeys typical patterns in Ce compounds \cite{els97}.
In heavy--fermions (e.g. CeCu$_2$Si$_2$ \cite{schl88},
CeNi$_2$Ge$_2$ \cite{kru98}) the Ce--spin fluctuations yield an
additional contribution to the usual Korringa relaxation. In
intermediate valent compounds (e.g. CePd$_3$ \cite{schae83},
CeOs$_2$ \cite{schl86}) the Korringa relaxation is strongly reduced
below $T^*$ because of the reduction of the density of states near
$E_F$. It is important to note that even a partial reduction of the
density of states strongly effects on the Gd$^{3+}$--relaxation,
whereas the electric resistance still exhibits metallic behavior as
long as $N(E_F) \not= 0$.

In CeNiSn both spin fluctuations and reduced density of states
influence the Gd--ESR. We will see that the ESR results in pure
CeNiSn closely agree with those of NMR experiments, and we present
a detailed and systematic investigation of the change of the gap in
the electronic density of states when doping Pt or Co on the Ni
sites. We find a transition to a purely metallic behavior in
Co--doped compounds and to a heavy--fermion behavior for the Pt
doped alloys. Hence, in the present work we could support and
extend the results of former investigations on CeNiSn by performing
ESR measurements at a wide temperature range from 1.6 to 120K in
high quality polycrystalline samples.

\section{Sample preparation and experimental setup}
Polycrystalline samples of $\text{Ce}_{0.99}\text{Gd}_{0.01}
\text{Ni}_{1-x}\text{Co}_x\text{Sn}$ and
$\text{Ce}_{0.99}\text{Gd}_{0.01}\text{Ni}_{1-y}\text{Pt}_y
\text{Sn}$ were melted together stoichiometrically from the
elements with a purity better than 99.9\% in an argon--arc furnace
and annealed for 5 days at 1073K. X--ray diffraction confirmed the
proper $\text{Co}_2\text{Si}$--structure and did not reveal any
parasitic impurity phases. Because of the anisotropy of the
paramagnetic susceptibility, which shows a maximum in the
crystallographic a--axis, it was possible to orient the samples:
The polycrystals were powdered to nearly single crystalline grains
by a mortar, embeded in liquefied paraffin, and oriented along the
a-axis within a static magnetic field of 17 kOe. The ESR
measurements were performed with a Bruker ELEXSYS spectrometer at
X-band frequencies ($\nu \approx 9$ GHz). For cooling the sample, a
continuous--flow helium cryostat (ESR900, Oxford Instruments) was
used for temperatures above 4 K and a cold--finger helium--bath
cryostat for temperatures below 4 K.

\section{Results}
\subsection{ESR spectra}
Electron--spin resonance probes the absorbed power $P_{\text{abs}}$
of a transversal magnetic microwave field with frequency $\nu$ as a
function of the static magnetic field $\boldmath{H}$. To improve
the signal--to--noise ratio, a lock--in technique is used by
modulating the static field, which yields the derivative of the
resonance signal $dP_{\text{abs}}/dH$.

\begin{figure}
\centering
\includegraphics[angle=-90,width=8.6cm,clip]{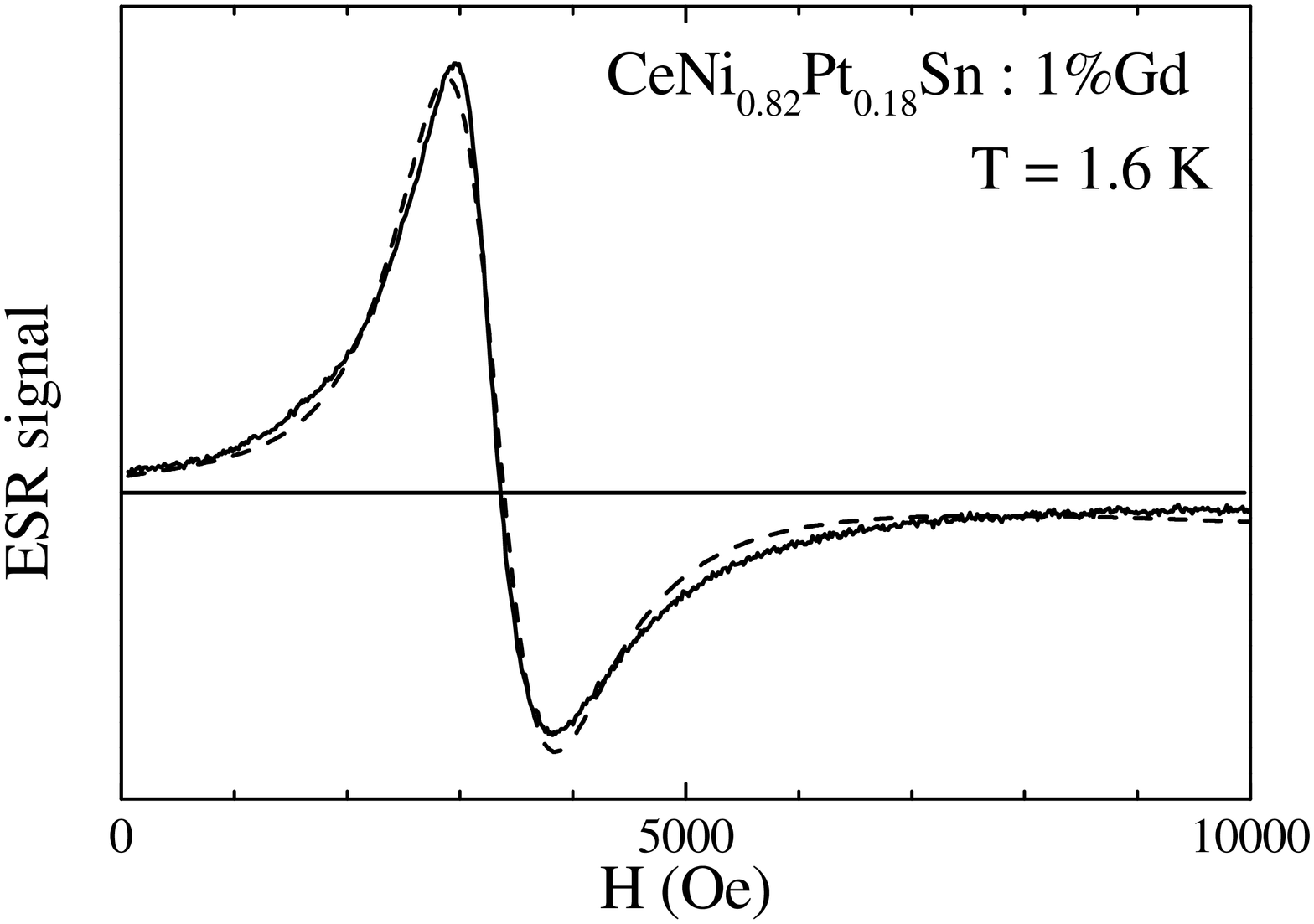}
\caption{Typical Gd$^{3+}$ ESR spectrum in
CeNi$_{0.82}$Pt$_{0.18}$Sn:1\% Gd (solid line). The result of a
fit, using equation (\ref{dyson}) is shown as dashed line.}
\label{spectrum}
\end{figure}

All compounds ($0\le x \le 0.1$ and $0\le y \le 0.2$) show a single
resonance line at about $g=2$. A representative result is shown in
fig.\ \ref{spectrum}\ as solid line. All spectra were fitted using
a Dysonian shape \cite{dys55}, given by
\begin{equation}
\frac{d}{dH} P_{\text{abs}} \propto \frac{d}{dH} \left(
\frac{\Delta H + \alpha \left( H-H_{\text{res}}\right)}{\left(
H-H_{\text{res}} \right) ^2+ \Delta H^2} +
\frac{\Delta H + \alpha \left( H+H_{\text{res}}\right)}{\left(
H+H_{\text{res}} \right) ^2+ \Delta H^2} \right)
\label{dyson},
\end{equation}
where $H_{\text{res}}$ denotes the resonance field and $\Delta H$
the half linewidth at half of the maximum absorption. In equation
(\ref{dyson}) a Lorentzian line ($\alpha=0$) is modified by the
influence of the skin effect, which in metals leads to a mixture of
real part ( = dispersion $\chi^{\prime}_{\text{Gd}}$) and imaginary
part ( = absorption $\chi^{\prime\prime}_{\text{Gd}}$) of the
dynamic susceptibility $\chi_{\text{Gd}}=\chi^{\prime}_{\text{Gd}}
+ i \chi^{\prime\prime}_{\text{Gd}}$. The
dispersion--to--absorption ratio $\alpha$ is used as fit parameter
$0 \le \alpha \le 1$. The resonance at $-H_{\text{res}}$ has to be
included, because $\Delta H$ is in the order of $H_{\text{res}}$.
The result of a representative fit is shown in Fig.\
\ref{spectrum}\ as dashed line.

To understand the origin of the spectra and their dependencies on
temperature and orientation, we have to look at the spin
Hamiltonian for the Gd$^{3+}$ probe in a metallic uniaxial
environment, given by \cite{bar81}:
\begin{equation}
{\cal H} =\mu_B \bbox{H} g \bbox{S} + \frac{1}{3}b_2^0
\left[3S_z-S\left(S+1\right)\right] + J_{\text{Gd}}\bbox{S}
\cdot \bbox{\sigma}. \label{hamiltonian}
\end{equation}
The first term describes the Zeeman interaction of the Gd spin
$\bbox{S}$ with the static magnetic field $\bbox{H}$ ($\mu_B$ is
the Bohr magneton and $g$ the gyromagnetic tensor, which is assumed
to be isotropic, with $g\approx 2$). The microwave with frequency
$\nu$ induces dipolar transitions between the equidistantly
splitted Zeeman levels yielding a single resonance line at $h \nu =
g \mu_B H$. With respect to the crystal symmetry, the
crystal--electric field (CEF) at the Gd--place is assumed to have
uniaxial character (parameter $b_2^0$), as described by the second
term. This leads to an orientation--dependent splitting of the
single spectrum into seven lines due to the Gd--spin S=7/2. The
third part of the Hamiltonian is the exchange interaction between
the Gd$^{3+}$ ion and conduction electrons with spin density
$\bbox{\sigma}$, where $J_{\text{Gd}}$ denotes the exchange
integral. If the exchange interaction is large compared to the
crystal--field splitting, the different dipolar transitions are
strongly coupled to each other, and the spectrum is exchange
narrowed into a single ESR line with orientation dependent
linewidth and resonance field. For a closer discussion of these
effects refer to \cite{kru98}. The spectra of all samples show only
a weak dependence on the orientation. The analysis of the spectra
following ref. \cite{kru98} results in low values for the uniaxial
CEF--parameter $0.3 \text{ GHz} \lesssim b_2^0 \lesssim 0.7$ GHz,
compared to the microwave frequency $\nu = 9$ GHz. Hence, the
influence of the crystal field can be neglected for all further
discussion.

\begin{figure}[t]
\centering
\includegraphics[width=8.6cm,clip]{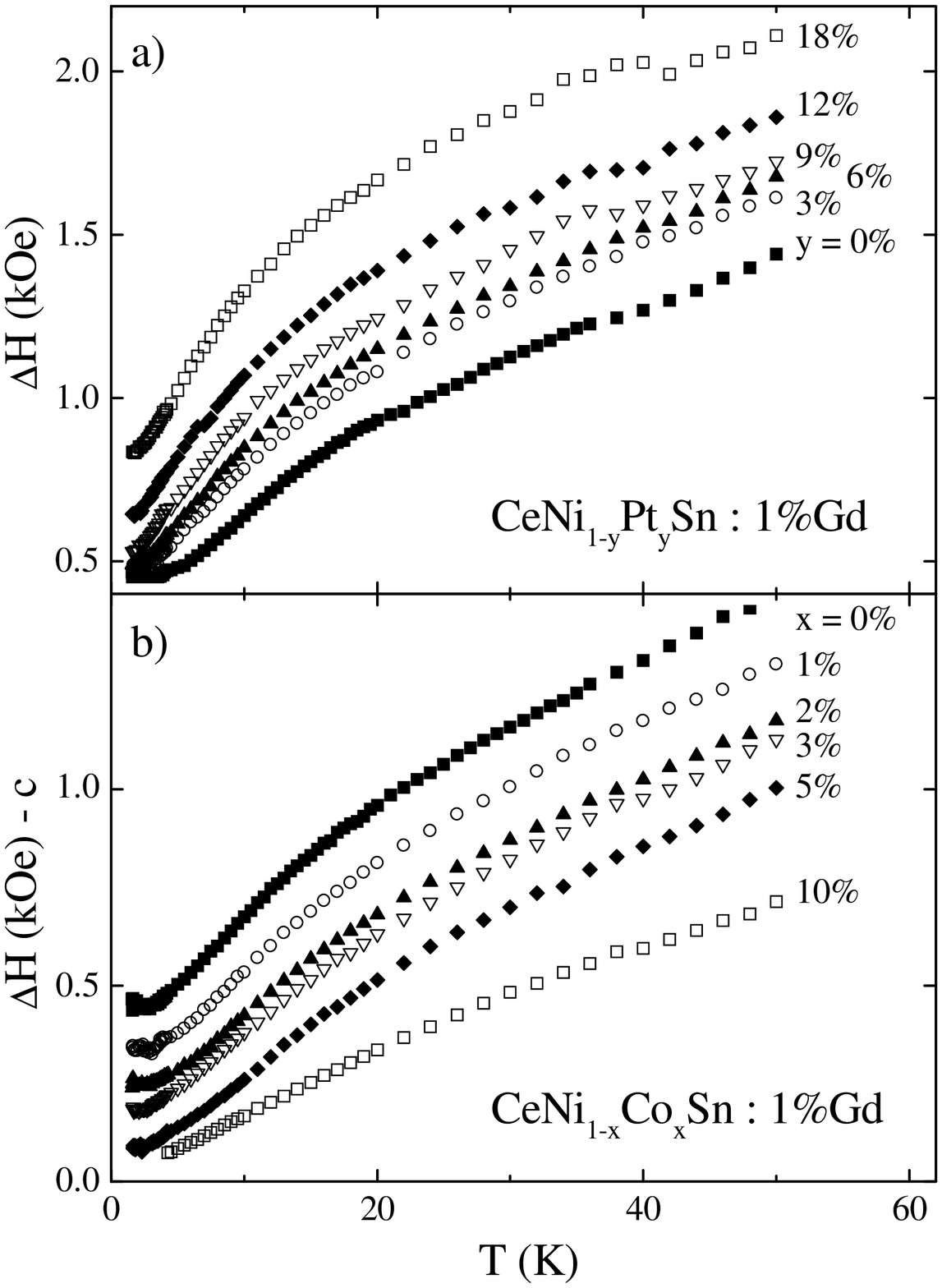}
\caption{Linewidth $\Delta H$ as a function of temperature T for
different concentrations of Pt (top) and Co (bottom). For better
clarity the curves in the lower frame are shifted to each other by
0.1 (x=1\%) to 0.5 (x=10\%).}
\label{dht}
\end{figure}

\subsection{Temperature dependence of ESR linewidth}
The temperature dependency of the linewidth was measured for
samples in the concentration range of interest. Fig.\ \ref{dht}a
shows the results for the Pt doped samples,
CeNi$_{1-y}$Pt$_y$Sn:1\% Gd for Pt concentrations $0 \le y \le
0.18$, Fig.\ \ref{dht}b the Co doped samples with $0 \le x \le
0.1$. Looking at the pattern of pure CeNiSn:Gd (Fig.\ \ref{dht}a:
$y=0$; Fig.\ \ref{dht}b: $x=0$), we recognize almost a Korringa
behavior with a slope $b$ of about 18 Oe/K for temperatures above
25K. At lower temperatures the slope increases yielding a gradient
$b\approx 34$ Oe/K for $5\text{ K} \le T \le 15\text{ K}$. At
lowest temperatures, $T<3$ K, the linewidth becomes almost
constant.

Focusing on the CeNi$_{1-x}$Co$_x$Sn compounds (Fig.\ \ref{dht}b),
we recognize, that the high temperature Korringa slope decreases
with increasing Co--concentration and also the transition region to
an enhanced slope shifts to higher temperatures. The sample with
10\% Co shows almost a pure Korringa behavior over the whole
temperature range. Substituting Pt for Ni (Fig.\ \ref{dht}a)
enhances the slope between 5 K and 15 K and shifts the transition
region to lower temperatures. At 18\% Pt the slope of $\Delta H$
continuously increases towards low temperatures, which is typical
for heavy--fermion systems with a rather low characteristic
temperature $T^*$ \cite{els97}.

\begin{figure}[t]
\centering
\includegraphics[angle=-90,width=8.6cm,clip]{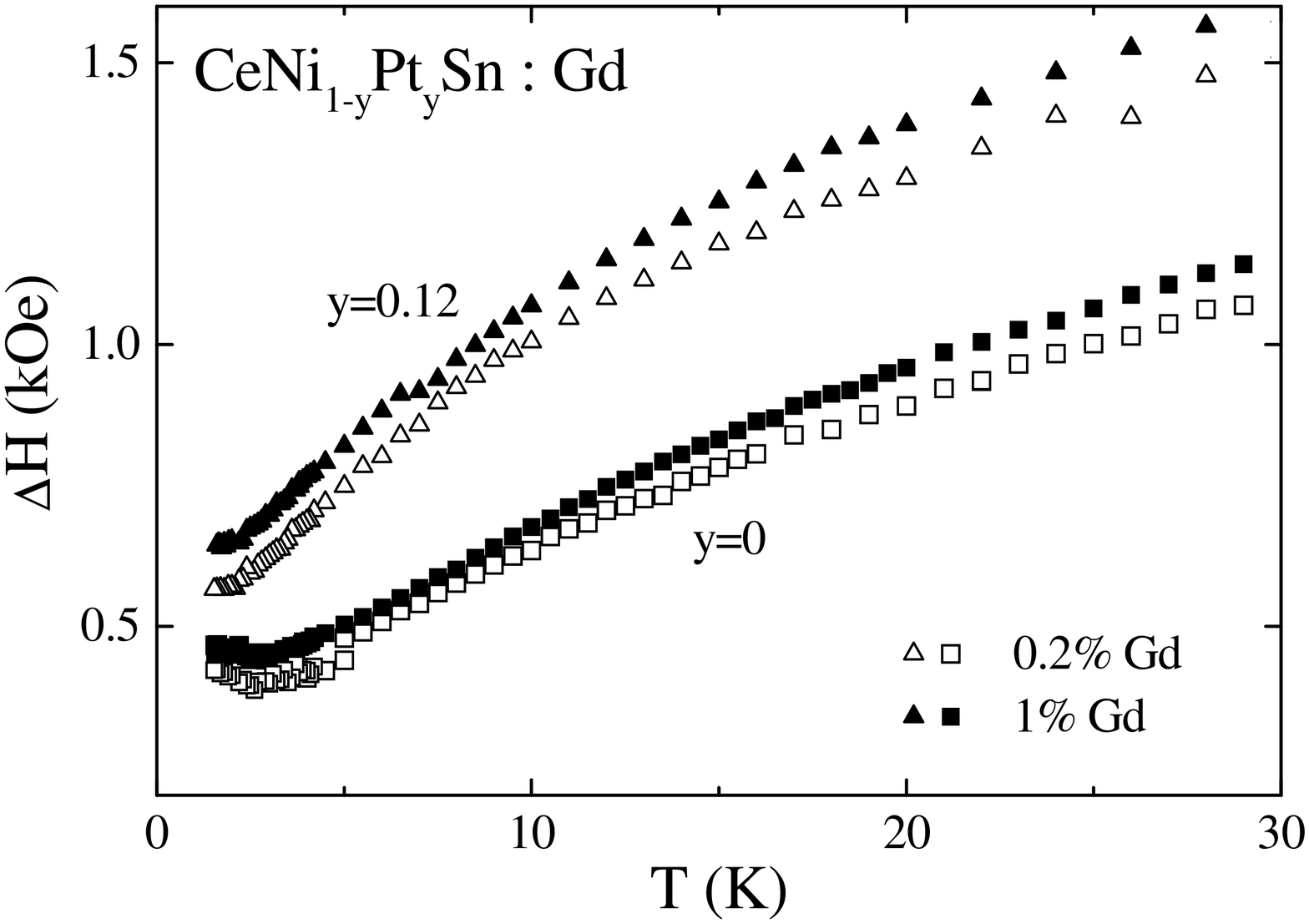}
\caption{Linewidth $\Delta H$ as a function of
temperature T for Gd--concentrations of 1\% (solid symbols) and
0.2\% (open symbols) for pure CeNiSn:Gd ($\Box$) and
CeNi$_{0.88}$Pt$_{0.12}$Sn:Gd ($\triangle$).}
\label{gd-vergleich}
\end{figure}

To determine the influence of Gd upon our results, we compared
samples with 0.2\% and 1\% Gd substituted at the Ce places. The
temperature dependence of the linewidth for these compounds is
shown in Fig.\ \ref{gd-vergleich} and clearly demonstrates that Gd
does not influence the linewidth. So we can exclude effects like
Gd--Gd--interaction or bottleneck effects, which are discussed in
Ref.\ \cite{col87}.

\begin{figure}[t]
\centering
\includegraphics[width=8.6cm,clip]{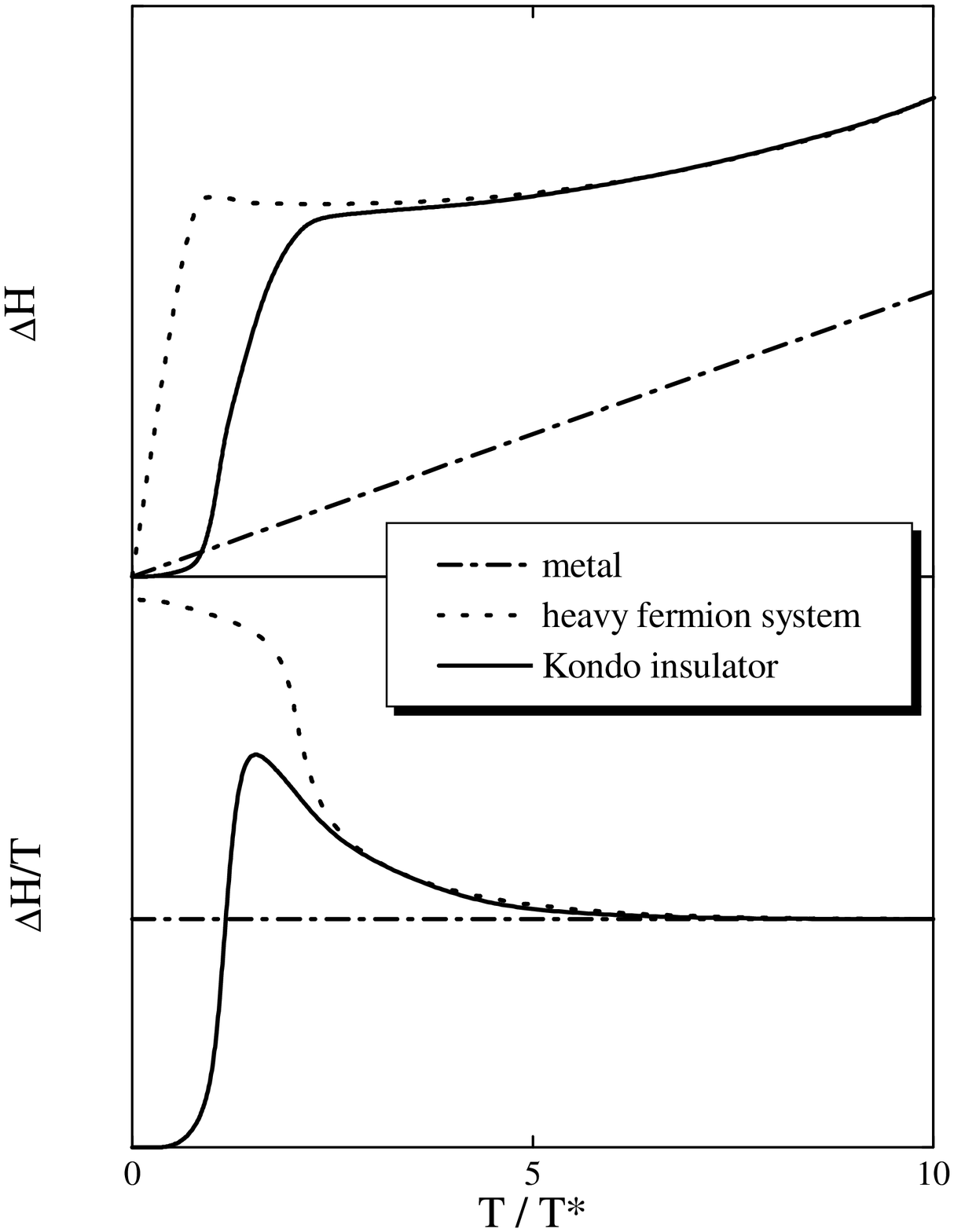}
\caption{Simulation of the ESR linewidth
of a metal, a heavy--fermion system and a
Kondo insulator.}
\label{modell}
\end{figure}

\section{Discussion}
Generally the ESR linewidth $\Delta H$ measures the transversal
relaxation rate $1/T_2$. Because of very short correlation times,
the longitudinal or spin--lattice relaxation rate $1/T_1$ equals
$1/T_2$ in metals \cite{bar81}. In heavy fermions the spin--lattice
relaxation of the Gd--spins is governed by two mechanism
\cite{col87,els97}:
\begin{equation}
\frac{1}{T_1} \propto \Delta H=\Delta H_{\text{K}} +
\Delta H_{\text{RKKY}}
\label{dh}
\end{equation}
The Korringa relaxation $\Delta H_{\text{K}}$, which is usually
observed in metals, is caused by the third term in equation
(\ref{hamiltonian}). The conduction electrons scatter at the
Gd--spin and transfer the energy to the lattice. This contribution
is proportional to the temperature T (dash--dotted curve in the
upper frame of figure \ref{modell}) and the square of the
electronic density of states at the Fermi level
$N_{\text{CE}}^2\left(E_F\right)$:
\begin{equation}
\Delta H_K \propto \left< J_{\text{Gd}}^2\left( q \right)\right>
\cdot N_{\text{CE}}^2\left( E_F \right) \cdot T := b \cdot T,
\label{korringa}
\end{equation}
where $\left< J_{\text{Gd}}^2\left( q \right)\right>$ is the
exchange integral averaged over the momentum transfer $q$ from the
scattering of the conduction electrons at the Gd--spin.

The second term $\Delta H_{\text{RKKY}}$ results from spin
fluctuations of the Ce--4f moments according to the
fluctuation--dissipation theorem. These are transferred to the
Gd--spin via RKKY interactions:
\begin{equation}
\Delta H_{\text{RKKY}} \propto T \chi_{\text{Ce}}^0 \tau \sum_i
\lambda_{\text{CeGd}}^2 \left(R_i \right), \label{rkky}
\end{equation}
where $\lambda_{\text{CeGd}}$ is the RKKY coupling to the Ce atoms
at a distance $R_i$. $\chi_{\text{Ce}}^0$ describes the static Ce
susceptibility, which in ideal heavy--fermion systems follows a
Curie--Weiss law $\chi_{\text{Ce}}^0 \propto \left(
T+\Theta\right)^{-1}$ with $\Theta=\sqrt{2} T^*$ \cite{cal91} at
high temperatures $T>T^*$ and saturates as a large Pauli--like
susceptibility of heavy quasiparticles at $T\ll T^*$. For $T
\ll T^*$ the spin--correlation time $\tau$ of the Ce--spins is
determined by the characteristic temperature according to
$1/\tau=k_BT^*/h$. At higher temperatures $T > T^*$ it shows a
behavior like $1/\tau
\propto \sqrt{T}$ \cite{cox85}. So, for $T < T^*$ we expect an
additional, strong linear increase of the linewidth $\Delta H$ with
the temperature (dotted curve in the upper frame of figure
\ref{modell}).

In contrast to the typical heavy--fermion pattern, for CeNiSn the
slope of $\Delta H (T)$ again decreases at lowest temperatures and
approaches zero. A similar behavior was observed in the
intermediate valent compounds CePd$_3$ and CeOs$_2$, where the
Korringa slope of the Gd linewidth was about 10 times smaller at
low temperatures than the Korringa relaxation $b$ in the respective
reference compounds YPd$_3$ and LaOs$_2$ \cite{schae83,schl86}.
This was explained in terms of a reduced density of states of the
conduction electrons within a regime $k_B T^*$ at the Fermi energy.
For a quantitative description $\Delta H_K$ was calculated with an
energy dependent density of states. The contribution $\Delta
H_{\text{RKKY}}$ vanished because of the delocalization of the
f--electrons in intermediate valent systems.

In CeNiSn, however, the contribution of the spin--fluctuations
cannot be neglected. Therefore a quantitative description would be
much more complicated. Here we confine ourselves to a qualitative
argumentation and to comparisons with results of other experimental
methods.

The gradual development of a pseudogap in the density of states of
both the conduction electrons and the 4f--electrons below a
characteristic temperature is a generic feature, which follows from
the coherence of a Kondo lattice \cite{gre91}. If the pseudogap
develops just at the Fermi level, it should reduce the relaxation
rate due to $\Delta H_K$ as well as $\Delta H_{\text{RKKY}}$, which
both depend on $N ( E_F)$. Therefore one expects an S--shaped
temperature dependence, which is shown as solid curve in the upper
frame of figure \ref{modell}. The excellent agreement of $\Delta H
(T)$ of the measured curve of CeNiSn, shown in Fig.\ \ref{dht},
with the model of a Kondo insulator as shown in Fig.\ \ref{modell}
is clearly evident.

For further discussion and comparison with NMR, susceptibility and
specific--heat experiments, we choose the $\Delta H/T$--plot, which
better pronounces the effect of the pseudogap. As can be seen from
eq.\ (\ref{dh} -- \ref{rkky}) $\Delta H / T$ corresponds to the
dynamic susceptibility plus a constant effect due to the Korringa
slope b. The theoretical expectations are shown in the lower frame
of figure \ref{modell}. Whereas normal metals reveal a constant
behavior over the whole temperature range and heavy fermions
saturate at a high value at low temperatures, Kondo insulators
exhibit a maximum in $\Delta H/T$ at a temperature
$T_{\text{max}}$, which is a rough measure for the width of the
gap.

Before we apply this plot on our experimental linewidth data, we
have to subtract the residual linewidth $\Delta H_0$, which is
always present in real samples and which is caused by impurities and
grating division errors. To obtain $\Delta H_0$, one has to
extrapolate the linewidth data to zero temperature.

\begin{figure}[t]
\centering
\includegraphics[angle=-90,width=8.6cm,clip]{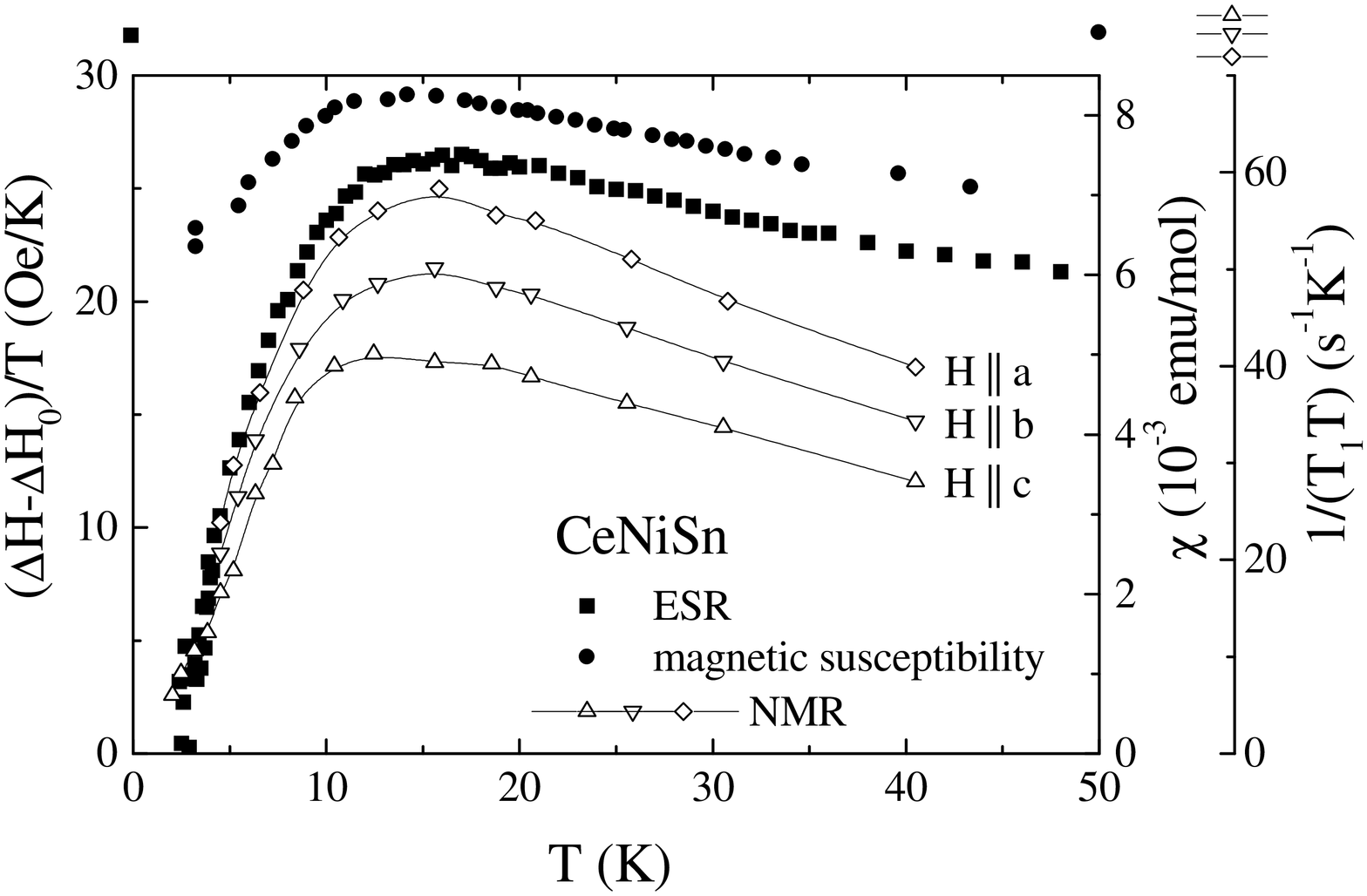}
\caption{Comparison of the results of ESR,
magnetic susceptibility
\protect\cite{tak96}, and NMR \protect\cite{oha95}.}
\label{verglcns}
\end{figure}

Figure \ref{verglcns} shows $(\Delta H - \Delta H_0)/T$, which
nicely resembles the expected pattern for a Kondo insulator. The
ESR results are compared to $^{119}$Sn--NMR and static
susceptibility $\chi^{\text{Ce}}_0$ results. The
$^{119}$Sn--nuclear spin relaxes via Fermi--contact interaction to
the conduction electron system. Therefore an analogous expression
to equation (\ref{dh}) holds for $(1/T_1)_{\text{NMR}}$. The NMR
data \cite{tak96} of $1/(T_1 \cdot T)$ show a similar behavior as
the ESR data. The orientation dependent measurements are very
similar to each other, in particular the position of the maximum,
which is a gauge for the gap, does not change with orientation.
These data also characterize CeNiSn as a Kondo insulator.

Macroscopic experimental methods like the magnetic susceptibility
$\chi$ \cite{oha95} prove the microscopic data: The decrease of
$\chi$ for low temperatures can be explained by a reduction of the
density of states at the Fermi level. As the correlation time
$\tau$ changes only slightly for $T<T^* \approx 20K$, $(\Delta
H-\Delta H_0)/T$ behaves similar to the susceptibility. At low
temperatures magnetic impurities may influence the magnetic
susceptibility significantly.

\begin{figure}[t]
\centering
\includegraphics[angle=-90,width=8.6cm,clip]{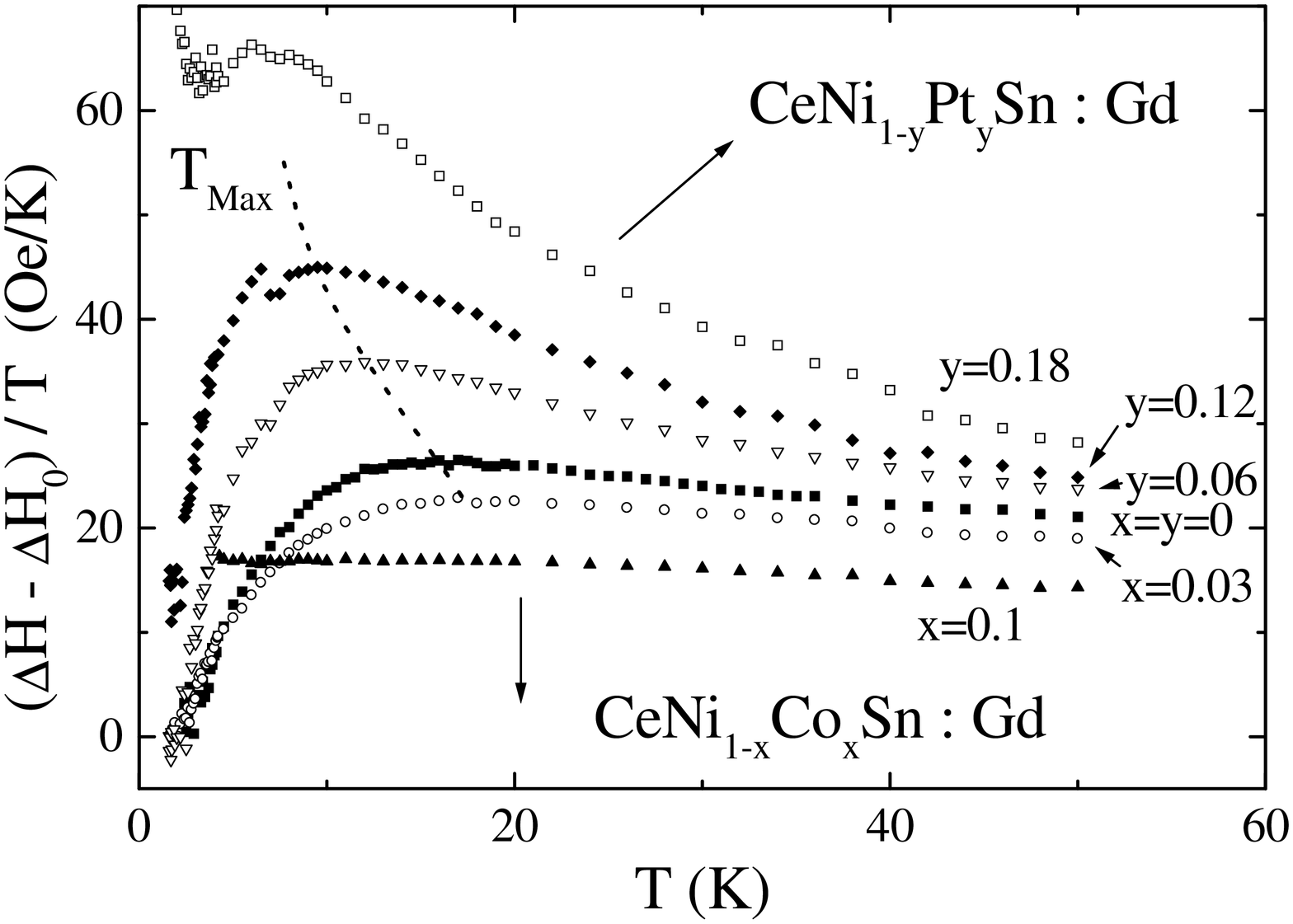}
\caption{$(\Delta H - \Delta H_0)/T$ as a
function of temperature
for different Co and Pt dopings.}
\label{dhtt}
\end{figure}

To examine the influence of the substitution of Ni by Co or Pt, we
use again the $(\Delta H - \Delta H_0)/T$ plot (Fig. \ref{dhtt}).
First we see a constant value for the compound with 10\% Co, which
behaves like a normal metal. The compound with a Pt concentration
of 18\% shows a constant high value for low temperatures, typical
for a heavy--fermion system. All other curves exhibit a behavior
like a Kondo insulator with a maximum at temperatures between 5 K
and 15 K. This maximum shifts to lower temperatures for increasing
Pt concentrations, and to higher temperatures for Co doping. As
described before, the maximum characterizes the width of the gap in
the density of states. However, the right extrapolation th $\Delta
H_0$ is sometimes quite difficult. To avoid a dependence on $\Delta
H_0$, we therefore introduced a characteristic temperature
$T_{\text{c}}$, which describes the temperature, at which the
bending or second differential on temperature changes from positiv
to negativ sign for increasing temperature. This turning point is
characteristic for the size of the gap, too.

\begin{figure}[t]
\centering
\includegraphics[angle=-90,width=8.6cm,clip]{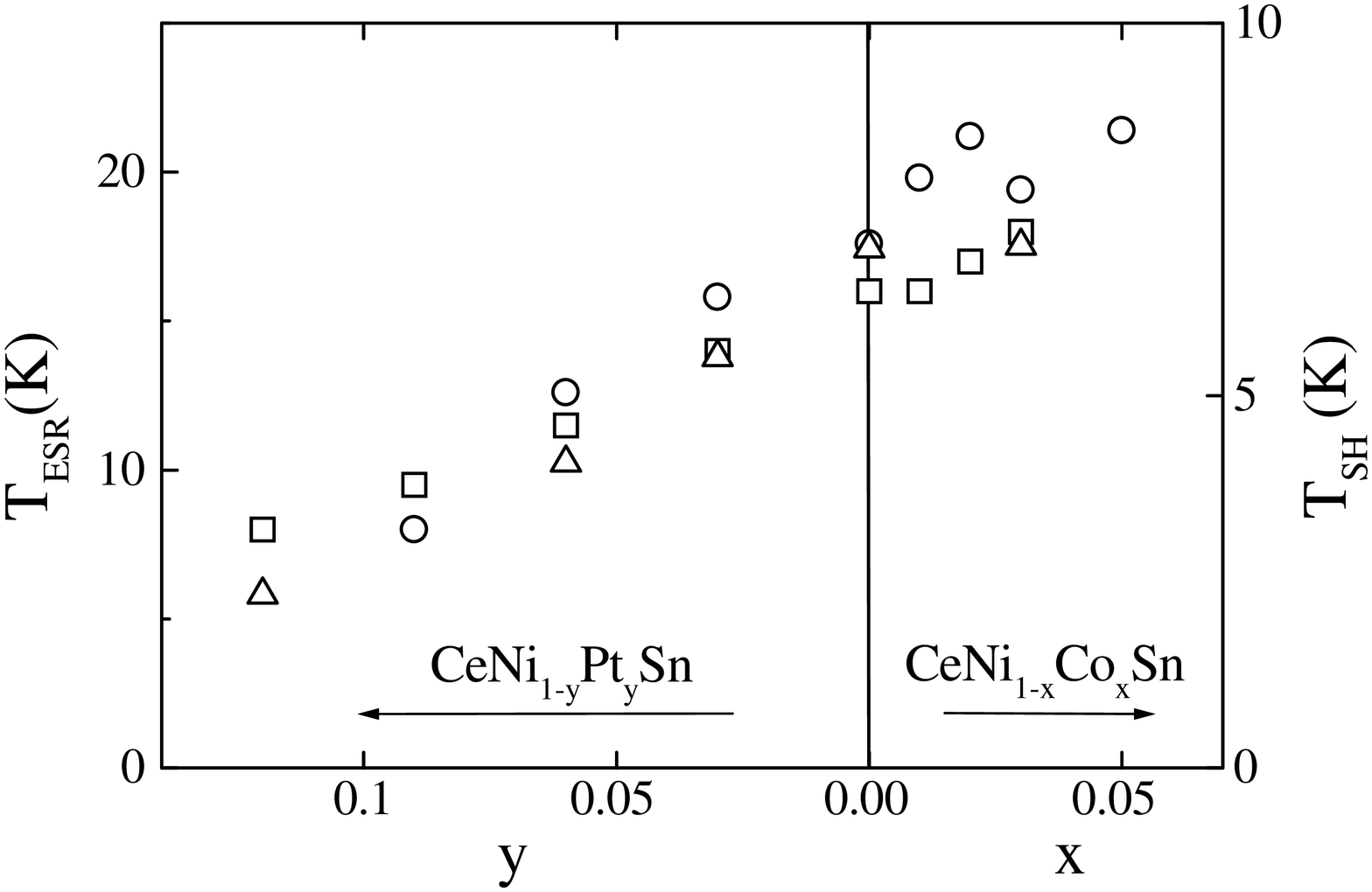}
\caption[Fig. 7]{Characteristic temperature ($\circ$,
left axis divided by 2) and maximum of $(\Delta H - \Delta H_0)/T$
($\Box$, left axis) of ESR measurements and maximum of $C_m/T$
($\triangle$, right axis) \protect\cite{nis96} for different
concentrations of Co (x) and Pt (y).}
\label{ueber_max}
\end{figure}

To determine the behavior of the gap with an increasing volume of
the elementary lattice cell, we substitute Ni by Pt. We clearly see
in fig.\ \ref{ueber_max}, that this leads to a decrease of this
characteristic temperature from 9 K at pure CeNiSn to 4 K with 9\%
Pt. Doping small amounts of Co into the samples leads to an
increase of the characteristic temperature, indicating an increase
of the gap. On the other hand, the behavior of the linewidth gets
more linearly. This points to the fact, that the density of states
at Fermi level is not zero, but becomes higher and higher for
increasing Co concentration. At about 10\% Co the gap is completely
filled, and the sample behaves like a metal.

Co substitution does not change the size of the elementary cell
volume. X--ray measurements revealed, that the lattice parameters
remain almost constant for doping up to 10\% Co on the Ni sites.
The most important effect of the Co seems to be the reduction of
the number of electrons. This leads to a destruction of the
Abrikosov--Suhl--resonance and so of the heavy--fermion behavior.
Figure \ref{ueber_max} also shows data of characteristic
temperatures for the gap, got by measurement of specific heat
\cite{nis96}. These data correspond very well to our ESR data.

\section{Conclusion}
We have shown that Gd--ESR nicely probes the coherence gap in Kondo
insulators. The S--shaped temperature dependence of the linewidth
in CeNiSn:Gd is comparable to the spin--lattice relaxation rate
$1/T_1$ of $^{119}$Sn--NMR, which both measure the
conduction--electron density of states at the Fermi energy. The
evolution of the gap by doping Pt or Co onto the Ni site agrees
with specific--heat experiments. Pt doping increases the cell
volume, hence decreases the hybridisation and therefore narrows the
gap, whereas Co changes the electron configuration, disturbs the
coherence, and the gap is gradually filled and smeared out.

\section*{Acknowledgments}
We are grateful to A. Brand, B. Mayer, S. Saladie and D. Vieweg for the
preparation of the samples and the X-ray-measurements. This work
was supported by the Bundesministerium f\"{u}r Bildung und Forschung
(BMBF) under Contract No. 13N6917/0.

\end{multicols}

\end{document}